\documentclass[fleqn,usenatbib]{mnras}

\usepackage{newtxtext,newtxmath}
\usepackage{graphicx}
\usepackage{xcolor}
\usepackage{subfigure}
\usepackage{ulem}
\usepackage{mathtools}
\usepackage{amsmath}
\usepackage{lastpage}
\newcommand{\omu}{'Oumuamua\,}
\newcommand{\lmb}{Lambertian\,}
\newcommand{\spc}{specular\,}
\newcommand{\ls}{Lommel-Seeliger\,}
\newcommand{\bks}{backscatter\,}
\newcommand{\degi}{^{\circ}}

\title[On the aspect ratio of \omu]{On the aspect ratio of \omu:\\ less elongated shape for irregular surface properties}
   
\author[Vazan \& Sari]{
Allona Vazan,$^{1,2}$\thanks{E-mail: allona.vazan@mail.huji.ac.il}
Re'em Sari,$^{1}$
\\
$^{1}$Racah Institute of Physics, The Hebrew University, Jerusalem 91904, Israel\\
$^{2}$Institute for Computational Science, Center for Theoretical Astrophysics \& Cosmology University of Zurich, Switzerland.} 

\date{}

\begin{document}
\label{firstpage}
\pagerange{\pageref{firstpage}--\pageref{lastpage}}
\maketitle

\begin{abstract}
The large brightness variation in the observed lightcurve of \omu is probably related to its shape, i.e., to the ratio between its longest axis and its shortest axis (aspect ratio).
Several approaches found the aspect ratio of \omu to be unusually elongated. 
Moreover, the spin axis orientation has to be almost perpendicular to the observer in order to obtain such an extreme lightcurve, a configuration which is unlikely.
However, interstellar \omu may have different surface properties than we know in our solar system. 
Therefore, in this work we widen the parameter space for surface properties beyond the asteroid-like models and study its effect on 'Oumuamua's lightcurve.
We calculate reflection from a rotating ellipsoidal object for four models: \lmb reflection, \spc reflection, single scattering diffusive and \bks.
We then calculate the probability to obtain a lightcurve ratio larger than the observed, as a function of the object's aspect ratio, assuming an isotopic spin orientation distribution.
We find the elongation of \omu to be less extreme for the \lmb and \spc reflection models.
Consequently, the probability to observe the lightcurve ratio of \omu given its unknown spin axis orientation is larger for
those models. 
We conclude that different surface reflection properties may suggest alternatives to the extreme shape of \omu, relieving the need for complicated formation scenario, extreme albedo variation, or unnatural origin. Although the models suggested here are for ideal ellipsoidal shape and ideal reflection method, the results emphasize the importance of surface properties for the derived aspect ratio.
\end{abstract}


\section{Introduction}
\omu, likely an interstellar object and the first one to be observed, exhibited a variation of about factor ten in its brightness\footnote{Brightness variation is found to range between 4.5-12 in different observations. While \cite{meech17} observed a lightcurve ratio (L$_{\rm max}$/L$_{\rm min}$) of more than 10 (2.5\,mag), the lightcurve ratio is between 6 and 9  (2$\pm$0.2\,mag) in \cite{jewitt17}, and as low as 4.5-8.2 (1.2-2.1\,mag) in \cite{bolin18}. Here we take brightness ratio of 10 as the standard.}  \citep[][etc.]{meech17,jewitt17,bolin18}.
This variation is linked mainly to the shape of \omu,
namely the ratio between its longest axis and its shortest axis (aspect ratio)  \citep{oumuamua19}.
The \omu aspect ratio was derived in several approaches \citep[e.g.,][]{meech17,bolin18,mcneill18} and is found to be extreme and much higher than the aspect ratio of objects in the solar system. 
Naively, one can expect the lightcurve magnitude variation between the brightest and the dimmest states (lightcurve ratio) to be equal to the aspect ratio. 
Indeed, the first work by \cite{meech17} fits the observed lightcurve ratio of 10 with a body of aspect ratio of 10, where the effect of the angle between the Sun and the observer (phase angle) is neglected. 

Previous estimates of the aspect ratio of \omu are based on reflection from asteroid-like objects: \cite{meech17} aspect ratio of 10 is based on \cite{detal94}, where the projection of the object is calculated assuming zero phase angle. 
\cite{bolin18} accounted for the actual phase angle and derived aspect ratio ranging from 4:1 to 10:1, based on the formulation of \cite{barucci82} for different models of asteroids. 
\cite{mashchenko19} finds aspect ratio of 8 to be most probable by using \ls reflection \citep{lumme81}. 
A somewhat lower aspect ratio of 6$\pm$1 is found by \cite{mcneill18} using the lightcurve inversion model of \cite{durech10}. This elongation is less than some other estimates, but still remarkable.

Nevertheless, elongation is expected to be greater than that. The above works assume the most favorable condition where the spin axis is perpendicular to the Sun -\omu- Earth (hereafter SOE) plane. If the spin axis is not perpendicular to the SOE plane the observed lightcurve ratio becomes smaller, i.e., the aspect ratio of the body has to be larger.
Thus, the resulting aspect ratios mentioned above are only lower limit for the real body's aspect ratio. The probability to observe an interstellar elongated object when its spin is exactly perpendicular to Earth is low. 
Since the orientation of \omu spin axis is unknown, there is higher probability that the \omu aspect ratio is larger than these lower limits, i.e. even more irregular \citep[e.g.,][]{mashchenko19,sirajleob19}.

It is well known that surface properties affect the light reflectance \citep{chandra60,lester79}.
The surface materials of some of the airless bodies in the solar system exhibit the opposition effect, which is a strong tendency to reflect light backward, to the source.
Other bodies are well described by the \ls law, a single scattering diffusive reflection \citep[e.g.,][]{muinonen15}.
However, irregular surface properties, different than the typical properties of solar system objects, might be more probable than the irregular shape that is derived for \omu.
Therefore, in this work we widen the parameter space for surface properties beyond the asteroid-like models.

In section~\ref{mthd} we calculate reflection from a rotating ellipsoid from basic principles.
We consider two extreme reflection cases: the perfect diffusive (\lmb), and the mirror (\spc). 
We calculate also single scattering diffusive (\ls) reflection, and \bks (projected area) reflection, which are relevant to objects in our solar system. In section~\ref{rslt} we derive the minimal aspect ratio of \omu for each method (obtained if the spin axis is perpendicular to the SOE), as well as the probability to get the observed lightcurve ratio (or larger) as function of the aspect ratio. We discuss the conditions for the suggested models and draw our conclusions in section~\ref{cncl}.

\section{Methods}\label{mthd}

We assume that \omu physical shape is a spheroid with axis ratio a:b:b , a$>$b, and that it has a uniform albedo. The object reflectance in each rotation angle is calculated ab-initio by integration over reflection from infinitesimal area elements on the surface of the ellipsoid. Spin axis orientation, and phase angle ($\Theta$) are naturally included. 

\subsection{Model geometry}

\begin{figure}
\centerline{\includegraphics[width=8.8cm]{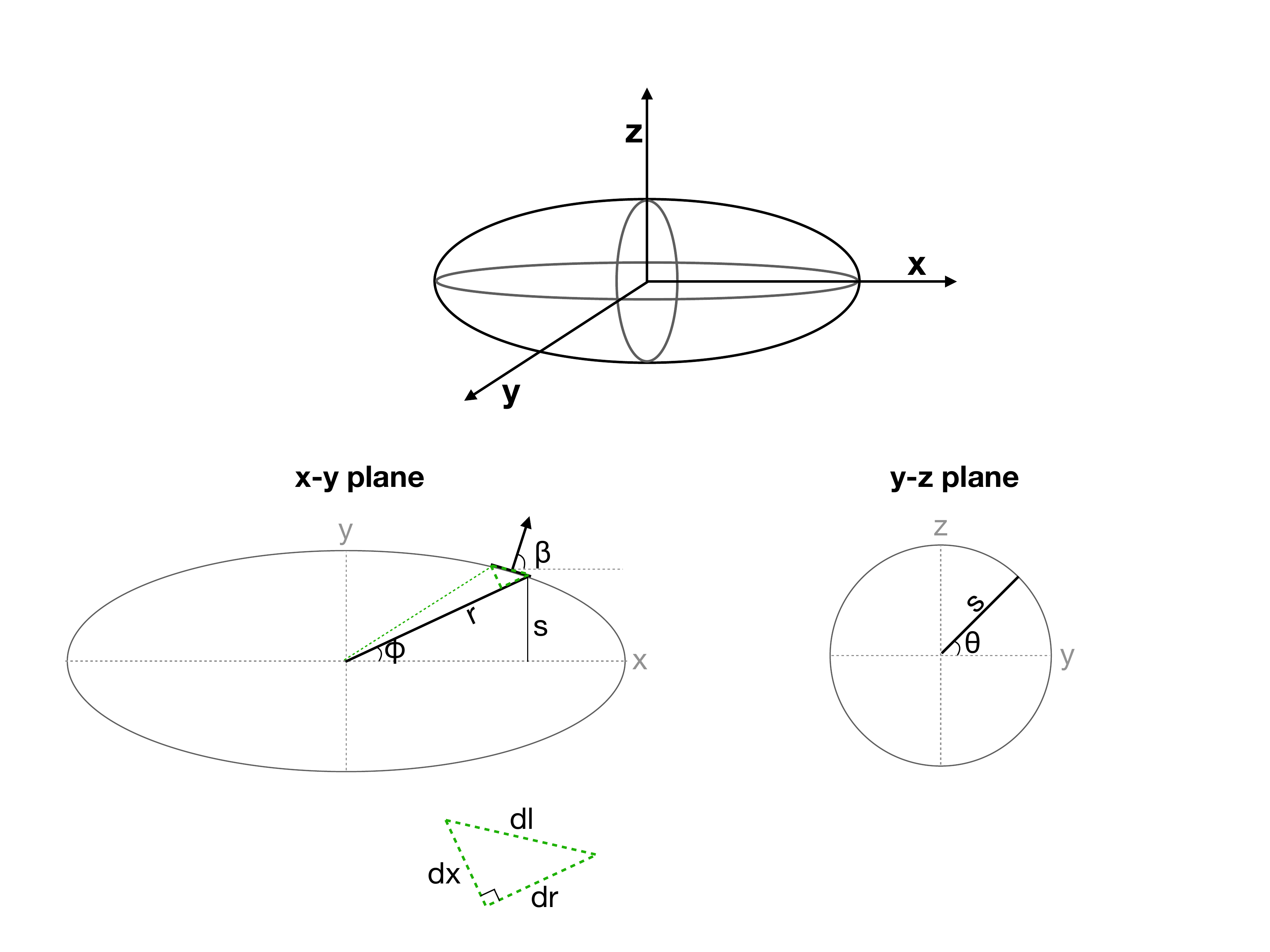}}
\caption{Geometry of the ellipsoid in our models. The dashed green rectangle is a zoom-in of the similar rectangle in the ellipse.
} \label{fig:cartoon0}
\end{figure}

 First we calculate the surface area elements of the ellipsoid.
 {The coordinates are fixed in the body's frame, as is shown in Fig.~\ref{fig:cartoon0}. The long axis of the ellipsoid is the x-axis, while the spin axis, parallel to the short axis of the ellipsoid is the z axis.} 
 For simplicity, the elements are build by multiplying intervals on ellipses in the x-y plane by intervals on circles in the y-z plane (see Fig.~\ref{fig:cartoon0}). 
The 2D ellipse (x-y) radius is described in respect to the azimuthal angle $\phi$ and the axes (a,b): 
\begin{equation}
     r(\phi)=\frac{a\cdot b}{\sqrt{a^2\sin^2\phi+b^2\cos^2\phi}} 
\end{equation}
In order to calculate the length interval ($dl$) in the x-y plane, we build a Pythagoras rectangle for each surface interval (see green dashed rectangle in Fig.~\ref{fig:cartoon0}). 
We use the change in ellipse radius with angle:
\begin{equation}
     dr(\phi)=\frac{a\cdot b}{2}\cdot \frac{\sin(2\phi)\cdot(b^2-a^2)}{\left(a^2\sin^2\phi+b^2\cos^2\phi\right)^{3/2}}\cdot d\phi
\end{equation}
and the radius interval of a circular object: $dx(\phi)=r(\phi)\cdot d\phi$.
The surface interval in x-y plane is then
\begin{equation}
    dl(\phi)=\sqrt{dr(\phi)^2+dx(\phi)^2}
\end{equation} 
The surface normal angle $\beta$ for each surface interval $dl$ is derived from $\phi$:
\begin{equation}
    \beta(\phi)=\phi+\arctan\frac{dr(\phi)}{dx(\phi)}
\end{equation}
In the y-z plane the surface intervals are of circular geometry (since b=c), in respect to the polar angle ($\theta$). The radius of each circle is determined by its location on the x-y plane ellipse: $s(\phi)=r(\phi)\cdot\sin\phi$. Thus, surface interval in the y-z plane is
\begin{equation}
    ds(\theta,\phi)=s(\phi)\cdot d\theta={r(\phi)\cdot\sin\phi}\cdot d\theta
\end{equation} 
The size of each surface element is calculated from the above geometry, and results in
\begin{equation}
    dA(\theta,\phi)=dl(\phi)\cdot ds(\theta,\phi)
\end{equation}
The normal of each surface area element is then determined by $\beta$ and $\theta$, forming a matrix of surface normals of the ellipsoid:
\begin{equation}
    \hat{A}_x=\cos\beta; \,\,\, \hat{A}_y=\sin\beta\,\cos\theta; \,\,\, \hat{A}_z=\sin\beta\,\sin\theta
\end{equation}

The Sun (projector) and the Earth (observer) are located in:
\begin{equation}
\begin{split}
   \hat{N_{\odot}}=[\cos\theta_{\odot}\cdot\cos\phi_{\odot}; \,\, \cos\theta_{\odot}\cdot\sin\phi_{\odot}; \,\,\sin\theta_{\odot}] \\
   \hat{N_{\oplus}}=[\cos\theta_{\oplus}\cdot\cos\phi_{\oplus}; \,\, \cos\theta_{\oplus}\cdot\sin\phi_{\oplus}; \,\,\sin\theta_{\oplus}] 
   \end{split}\label{eq:soe}
\end{equation}
These angles are related to the phase angle $\Theta$ by $\cos \Theta=\hat{N_{\odot}} \cdot \hat{N_{\oplus}}$.
The effect of the rotation axis orientation on the lightcurve is obtained by varying $\phi_{\odot},\phi_{\oplus}$. 
For \omu, where the phase angle is known, we consider only SOE configurations that are consistent with this known value. 
%; {namely, all the angles that satisfy $\Theta=\sphericalangle(\hat{N_{\odot}},\hat{N_{\oplus}})$.}

The observed brightness ratio depends also on the spin axis orientation in respect to the SOE plane.
Since the spin axis orientation of \omu is unknown, we vary this parameter. First, we assume that the spin axis orientation is perpendicular to the SOE (hereafter perpendicular spin). 
Perpendicular spin results in the largest lightcurve ratio for given SOE and object, and the results in Sec.~\ref{rslt1} are under this assumption. 
We relax this assumption in Sec.~\ref{rslt2}, when we allow for various spin axis orientations and calculate the probability to observe \omu lightcurve for different aspect ratios.

\subsection{Model assumptions}
(1) We assume the geometry of the \omu to be ellipsoid with a$>$b=c, i.e., an elongated shape. The elongated (cigar-like) shape is more likely than a flat (pancake-like) shape, both because it is more energetically stable and because it has a larger range of possible orientations \citep{belton18,katz18,oumuamua19}. We therefore ignore the pancake-like shape, although it was found to be more probable by \cite{mashchenko19}.
The assumption of $b=c$ is consistent with calculation by \cite{belton18}, that found $b=1.03c$ for a cigar shape. \\
(2) We take the SOE location and distances to be constant during rotation period. To make this assumption we compare the \omu rotational period time ($\sim8\,hr$) with the location change in this time. With an average velocity of $26\,km/s$ the change in location within a rotation period can be as much as $0.005\,AU$. This distance is small in comparison to \omu closet point to Earth ($0.16\,AU$).\\
(3) We assume a constant phase angle, because the SOE location is nearly constant during one rotation period. During the overall observation period of \omu, the phase angle changed between $19\degi - 27\degi$, and up to $24\degi$ for the mag$\,\geq\,$2.2 observation period \citep{jewitt17,bolin18,mcneill18,belton18}. Therefore we calculate here for $\Theta=2\pi/15=24\degi$, but also consider cases of $\Theta=\pi/9=20\degi$, to show the effect of phase angle on the results.\\
(4) We assume a uniform albedo for simplicity. \\
(5) We assume rotation only around the z-axis (see Fig.~\ref{fig:cartoon0}) from minimum energy consideration.\\ 
(6) We ignore tumbling for simplicity of the model. \omu photometry data suggests that it is tumbling \citep{fraser18,belton18}, as the lightcurve and rotation time change between different periods. These variations are of order 10\%.\\
(7) We ignore the difference in lightcurve ratio between October and November measurements \citep{belton18}.\\ 

\subsection{Surface reflection}

\begin{figure}
\centerline{\includegraphics[width=8.8cm]{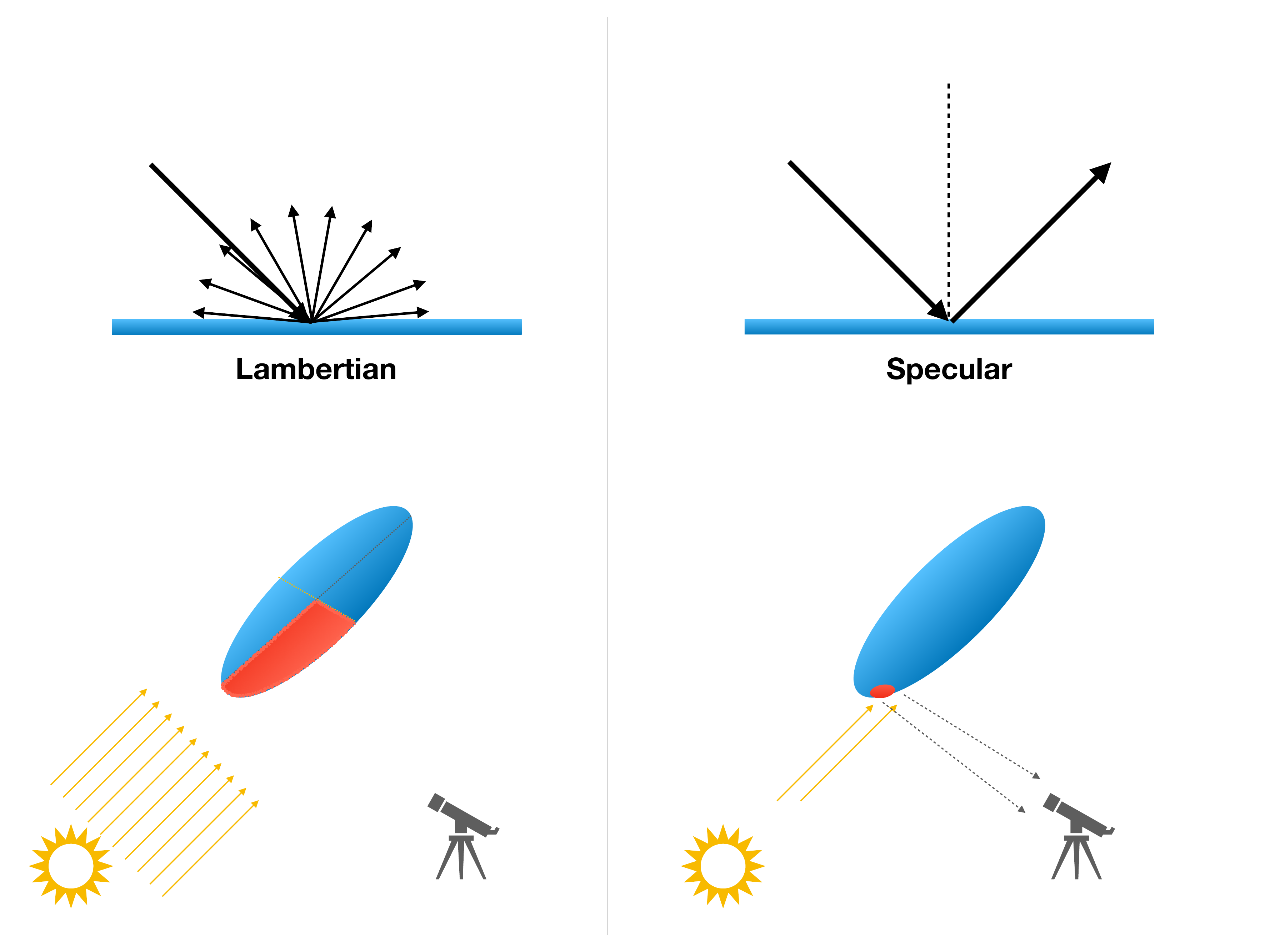}}
\caption{
An illustration of \lmb (left) and \spc (right) reflection.
Up: reflection from a plane unit surface area. Bottom: overall reflection from an ellipsoid.
The surface area on each ellipsoid that contributes the observed brightness appears in red.} \label{fig:cartoon}
\end{figure}

\underline{\lmb surface:}
If \omu has a matte (perfect diffusive, \lmb) surface, the apparent brightness from a surface element is the same for all observe angles of view (see Fig.~\ref{fig:cartoon} upper left). 
For a finite surface element the reflectance from \lmb surface is determined by the incident flux angle ($\cos\psi_i$), but also by the scattered flux angle ($\cos\psi_s$) which affects the angular size of this element \citep[e.g.,][]{durech10}. The flux from each infinitesimal surface element is then
\begin{equation}
    dF \propto dA \cdot \cos\psi_i \cdot \cos\psi_s
\end{equation}
where
\begin{equation}
    \cos\psi_i=\hat{dA}\cdot\hat{N_{\odot}}; \,\,\, \cos\psi_s=\hat{dA}\cdot\hat{N_{\oplus}}\label{eq:cos}
\end{equation}
The overall flux is obtained by integration of dF over all ellipsoidal surface elements that are both illuminated $(\cos\psi_i>0)$ and observable $(\cos\psi_s>0)$, as marked in red in the bottom left of Fig.~\ref{fig:cartoon}. \\

\underline{Specular reflection:}
Here we take \omu to be a perfectly polished ellipsoid, acting as a mirror (\spc reflection).
For a given ellipsoid position and orientation, reflection is received from a single point on the surface, where the surface normal ($\hat{dA}$) is in the same direction as the bisector of the SOE angle,  i.e. parallel to  $\hat{N_{\oplus}}+\hat{N_{\odot}}$, as is shown in the right panels of Fig.~\ref{fig:cartoon}. 
The reflected intensity is inversely proportional to the curvature at this point as more curved surface scatters in a wider solid angle.
Thus, for an axis symmetric body, the reflected flux is proportional to:
\begin{equation}
    F \propto \frac{dl(\phi_r)}{d\beta}\cdot \frac{r(\phi_r)\cdot\sin\phi_r}{\sin\beta}  
    \label{eq:spc}
\end{equation}
where $\phi_r$ is the coordinate of the point of reflection - the point where the surface normal is parallel to the SOE bisector.
For an ellipsoid, Eq.~\ref{eq:spc} can be calculated analytically:
\begin{equation}
    F(\phi_r) \propto \frac{a^2 b^2 \left(a^2\sin^2\phi_r+\frac{b^4}{a^2}\cos^2\phi_r\right)^2}{\left(a^2\sin^2\phi_r+b^2\cos^2\phi_r \right) \left(a^2b^2\sin^2\phi_r+b^4\cos^2\phi_r \right)} \label{eq:spc2}
\end{equation}
Note, that due to the axis symmetry of the body, the flux depends only on $\phi$ and not on $\theta$.
When we set the angle in Eq.~\ref{eq:spc2} to the extreme values of $\phi =0$ and $\phi=\pi/2$, which are obtained for perpendicular spin,
we find that the lightcurve ratio for ellipsoid with \spc surface is as high as $(a/b)^4$. 
Such a ratio is extreme - ellipsoid with $a/b=2$ produces a lightcurve ratio of 16! This is a result of the extreme assumptions of a perfect mirror surface and a perfect ellipsoidal shape. Any perturbation on the shape or in the surface smoothness will lower this ratio, i.e., the perfect \spc reflection provides the most extreme change of lightcurve ratio with aspect ratio.\\

\underline{Backscatter:}
The simplest case of reflection is the projected area law \citep[e.g.,][]{connelly84}.
When the Sun and the observer are in the same object-centric direction (zero phase angle) the area of projection of the ellipsoid toward Earth direction is the cross-section of the ellipsoid geometry. In this case the lightcurve max/min ratio is proportional to the aspect ratio of the ellipsoid. 
The projected area law is used to model \bks reflection, where radiation is reflected back toward the Sun. 
In a perfect \bks process all radiation is reflected back and the object is not seen if the phase angle is not zero. 
In a non-perfect \bks reflection most of the light is reflected back, but some light is reflected in other directions, with the intensity decreases as the angle from the Sun direction gets larger (actually the phase angle). The key point is that the intensity variation depends only on the phase angle, and is independent of the object orientation.
Therefore, since the phase angle is approximately constant\footnote{Phase angle variation is small during 'Oumuamua's rotation period, see model assumptions.} the observed lightcurve ratio is proportional to the projected area ratio.
The flux from a surface element is then:
\begin{equation}
    dF \propto dA \cdot \cos\psi_i
\end{equation}
The total flux is the sum over all surface elements with $\cos\psi_i>0$ and $\cos\psi_s>0$.\\

\underline{\ls model:}
Lightcurve inversion models are usually derived under the assumption that the light-scattering behaviour of asteroids can be described as a combination of single-scatter diffusive (\ls) and \lmb models \citep[e.g.,][]{durech10}. 
Moreover, the \ls model is a simple model that fits well the more detailed Hapke model for asteroids \citep{hapke02,hapke12,huang17}.
Therefore, we calculate also reflectance from \ls surface.
Here, the contribution of each surface area element is determined by the incident flux angle ($\psi_i$) and the scattered flux angle ($\psi_s$) via \citep[e.g.,][]{fairbairn05,durech10}: 
\begin{equation}
    dF \propto dA\cdot \frac{\cos\psi_i \cdot \cos\psi_s}{\cos\psi_i + \cos\psi_s}
\end{equation}
Also here, we sum over all surface elements with $\cos\psi_i>0$ and $\cos\psi_s>0$ to get the total flux.\\

For each of the surface reflection models we repeat the calculation for all rotation angles along one period ($2\pi$), for a given spin axis orientation.
By that we actually assume that the spin axis is always around the short axis (z direction in Fig.~\ref{fig:cartoon0}).
Then, we take the ratio of the maximum value to the minimum value of the intensity vector to be the lightcurve ratio.

\subsection{Lightcurve ratio probability}

For a given body shape and reflectance properties, the observed lightcurve ratio depends on the spin axis orientation. Since \omu spin axis orientation is unknown, we calculate the lightcurve ratio for all possible rotation axis orientations.
The spin axis orientation is defined by its azimutal ($\phi_s$) and polar ($\theta_s$) angles.
For each model we calculate the lightcurve ratio with a given aspect ratio for $10^6$ cases of spin axis orientations distributed isotropically.
Using these $10^6$ results for the lightcurve ratio, we calculate the probability that the lightcurve ratio is above the observed value.

The probability that we discuss here is the probability to observer a lightcurve ratio above the observed value, for a given surface properties given a random orientation of the spin direction.
This probability should not be interpreted as the probability that the object has such surface properties.
The \bks and \ls surface properties are more in line with the scattering properties of objects in our solar system than the \lmb or \spc reflection surfaces.

\section{Results}\label{rslt}

In section~\ref{rslt1} we calculate the minimal aspect ratio needed to obtain a certain lightcurve ratio, assuming perpendicular spin (i.e., the largest lightcurve ratio for a given aspect ratio). 
In section~\ref{rslt2} we calculate the probability to observe a lightcurve ratio equal or above the observed ratio, as a function of the aspect ratio of the body, assuming an isotropic spin distribution.

\subsection{Minimal aspect ratio}\label{rslt1}
For each of the models we calculate the reflection vs. rotation angle within one rotation period, and vary the aspect ratio in order to achieve a lightcurve ratio of 10. 
For zero phase angle ($\Theta=0$), the \bks reflection as well as the reflection by \ls model require aspect ratio of 10, as expected. When a phase angle of $\Theta=24\degi$ is applied the lightcurve ratio of 10 is obtained by aspect ratio of $\sim5.5$ for \bks reflection and $\sim5$ for \ls surface reflection.

However, if the surface is \lmb the minimal aspect ratio is as small as $\sim3.5$ with $\Theta=24\degi$, and of $\sim4$ with zero phase angle\footnote{Changing the phase angle to $\Theta=20\degi$ results in aspect ratio of 6 for \bks model, 3.6 for \lmb surface, and 5.4 for \ls surface.}. Specular surface requires aspect ratio of only $\sim1.8$ to provide 10 lightcurve ratio, and is independent of phase angle. 
These values, although calculated for extreme surface properties, are much lower than what was found in previous works. 
Thus, if \omu's surface reflection properties significantly differ from those of asteroids, its shape might be much less elongated.

\begin{figure}
\centerline{\includegraphics[width=8.8cm]{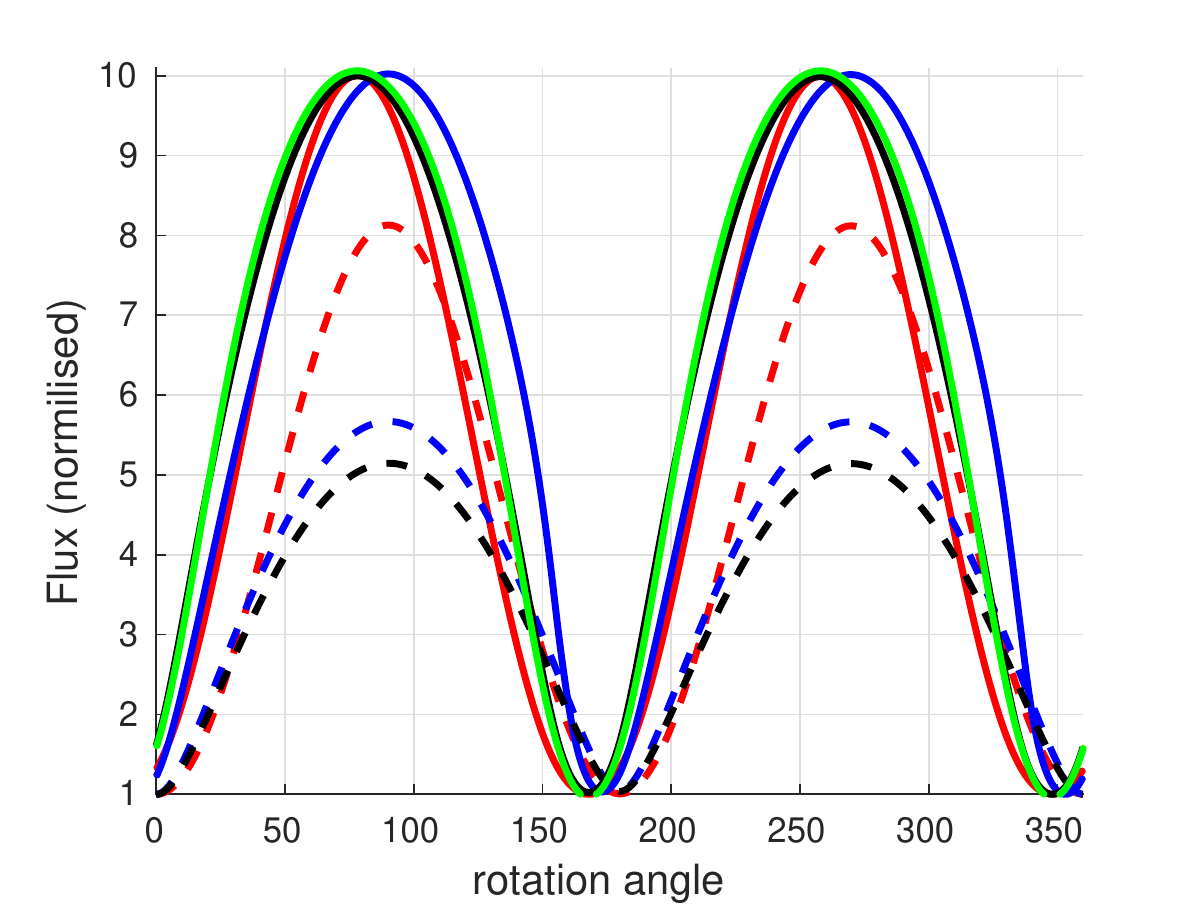}}
\caption{Calculated lightcurves for spinning ellipsoids with different surface reflection properties: \lmb (red), \spc (green), \bks (blue), and \ls (black). Solid curves are for phase angle of $\Theta=24\degi$, where the aspect ratio of each ellipsoid (3.5, 1.8, 5.5 and 5 respectively) where chosen to achieve the observed lightcurve ratio of 10. Dashed curves are for the same aspect ratios with $\Theta=0\degi$. Rotation spin axis is perpendicular to the SOE plane.
}\label{fig:orbit}
\end{figure}

In Fig.~\ref{fig:orbit} we show that for a finite phase angle ($\Theta=24\degi$) the lightcurve ratio is larger than for zero phase angle. 
The calculated lightcurves for all reflection models are shown in the figure. The solid curves are for the above aspect ratios to get brightness ratio of 10 with a phase angle of $\Theta=24\degi$. The dashed lines are for the same objects, but with $\Theta=0\degi$.

The effect of phase angle is not similar for different reflection methods.
The greatest effect of the phase angle is for \bks reflection and for the \ls reflection. 
For \lmb surface the phase angle effect is smaller. Although the phase angle changes the overlap between the incident flux area and the seen area (as is shown in Fig.~\ref{fig:cartoon}), the isotropic flux by the \lmb surface diminishes this effect. 
The \spc reflection amplitude is not affected by the phase angle, since for any phase angle there is a given surface element in the SOE plane that reflects the flux (as is shown in Fig.~\ref{fig:cartoon}). Therefore, the lightcurve is shifted, but the max/min ratio remains the same.

\begin{figure}
\centerline{\includegraphics[width=8.8cm]{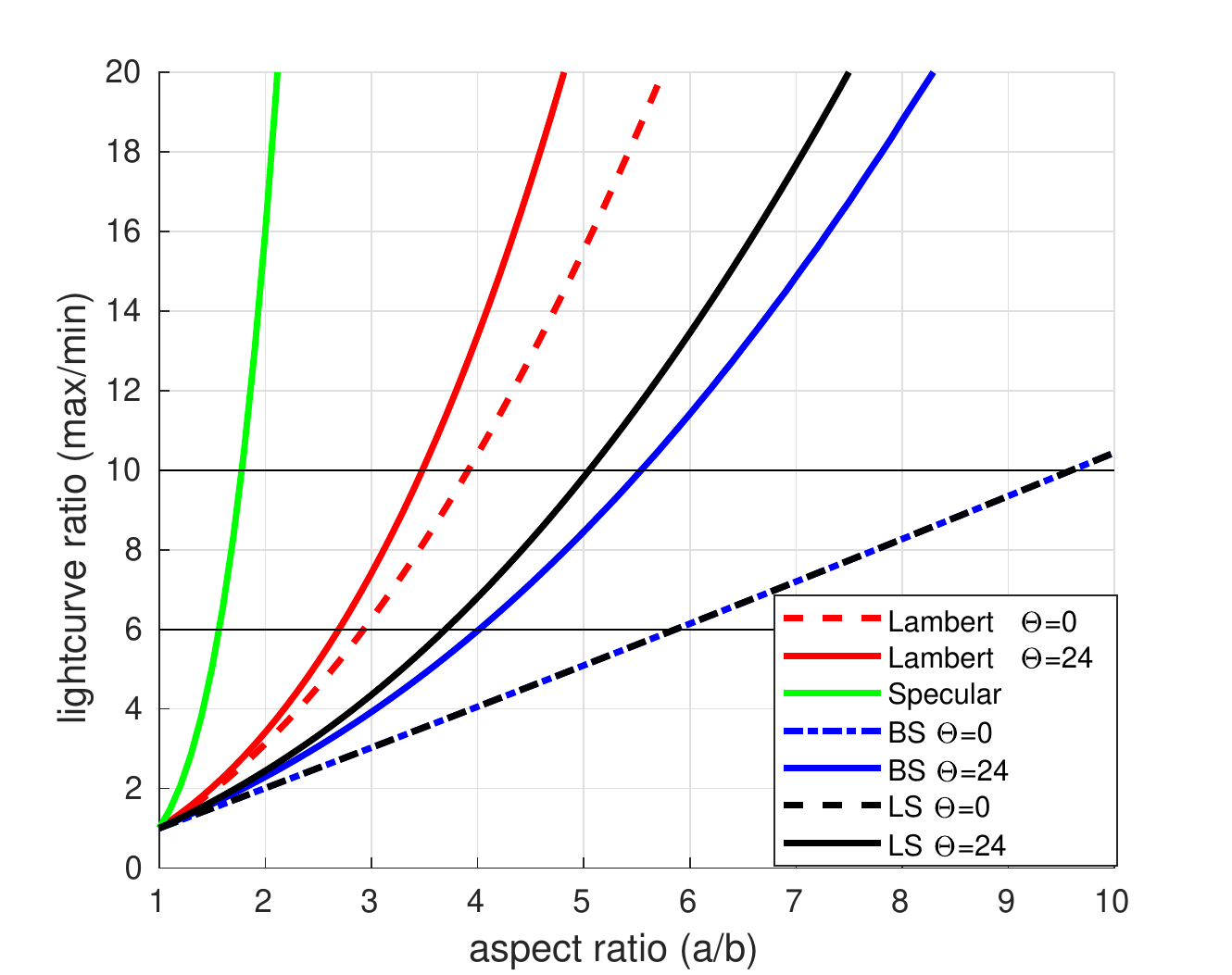}}
\caption{Lightcurve ratio as a function of the aspect ratio of ellipsoid (b=c), for different surface reflection conditions: \lmb   (red), \spc (green), \bks (blue), and \ls (black). Dashed curves are for phase angle $\Theta=0\degi$ and solid for $\Theta=24\degi$. The green curve for \spc reflection is independent of the phase angle. Rotation axis is perpendicular to the SOE plane for all cases.}\label{fig:maxmin}
\end{figure}

In Fig.~\ref{fig:maxmin} we show how the lightcurve ratio (max/min) changes as a function of the aspect ratio (a/b) of the object. We consider cases with phase angle contribution ($\Theta=24\degi$) and without ($\Theta=0\degi$). 
The observed lightcurve ratio of 10 \citep{meech17} and 6 \citep{jewitt17} are marked (horizontal lines). The intersection of the observed lightcurve with a model curve marks the minimum aspect ratio for this model. 
The dependence of the lightcurve ratio on the aspect ratio, i.e., the slope of the curve, differs between the surface models.
The \lmb surface and the \spc surface have steeper slopes and thus smaller aspect ratio to explain the \omu observations, in comparison to the other models.
As expected, the effect of phase angle, which was shown in Fig.~\ref{fig:orbit}, is greater as the aspect ratio increases for the same surface conditions.

Our calculation were done for a lightcurve ratio of 10. However, \omu brightness variation ranges between 4.5-12 in different observations. 
For a lightcurve ratio lower than 10 the aspect ratios get smaller, as is shown in Fig.~\ref{fig:maxmin}. 
For example, if we consider a lightcurve ratio of 6, the elongation of \omu with \lmb surface is only 2.7, while the \ls and \bks reflection models results in 3.7 and 4 respectively. 
For the perfect \spc reflection surface the aspect ratio is as small as 1.55.

\subsection{Probability to observe 'Oumuamua's lightcurve ratio}\label{rslt2}

\begin{figure*}
\subfigure{\includegraphics[width=8.4cm]{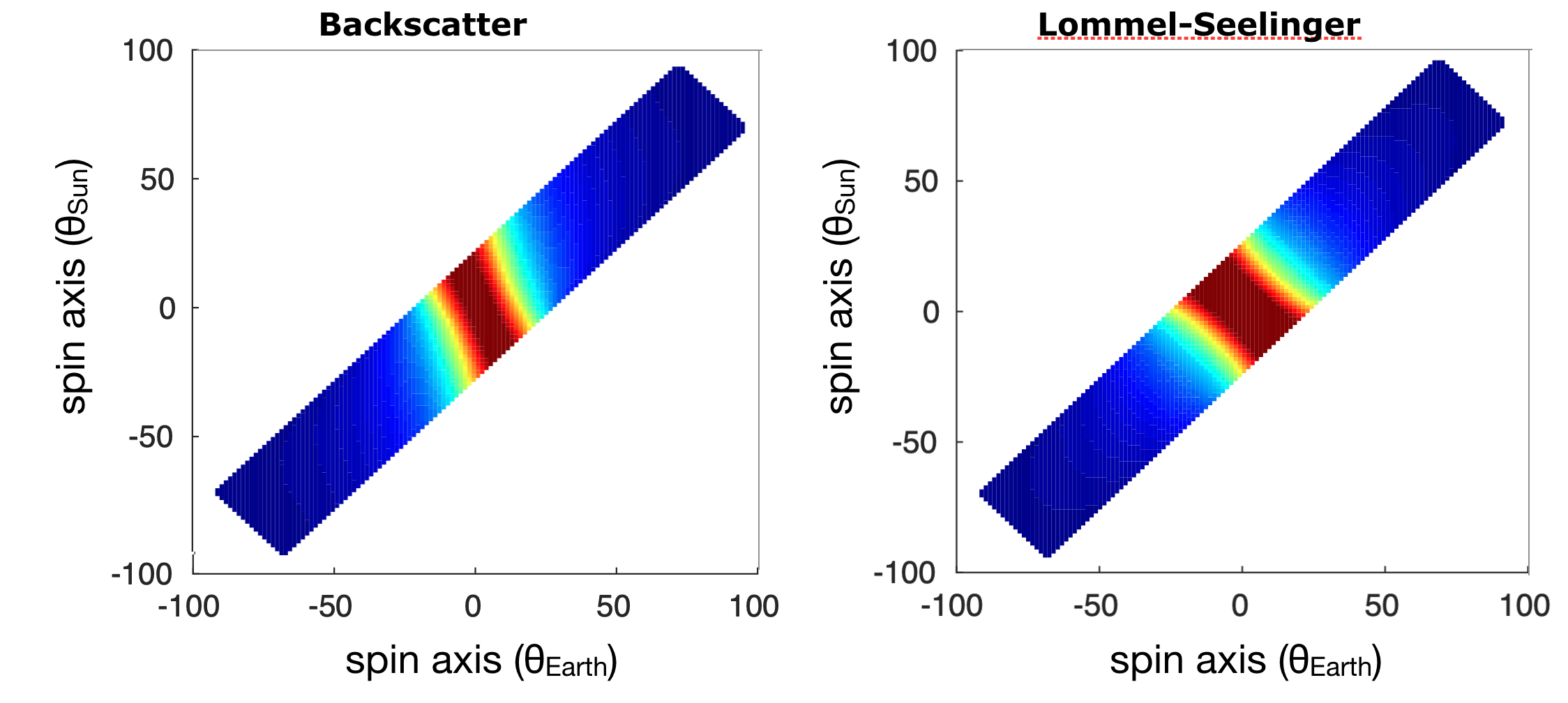}}
\subfigure{\includegraphics[width=9.0cm]{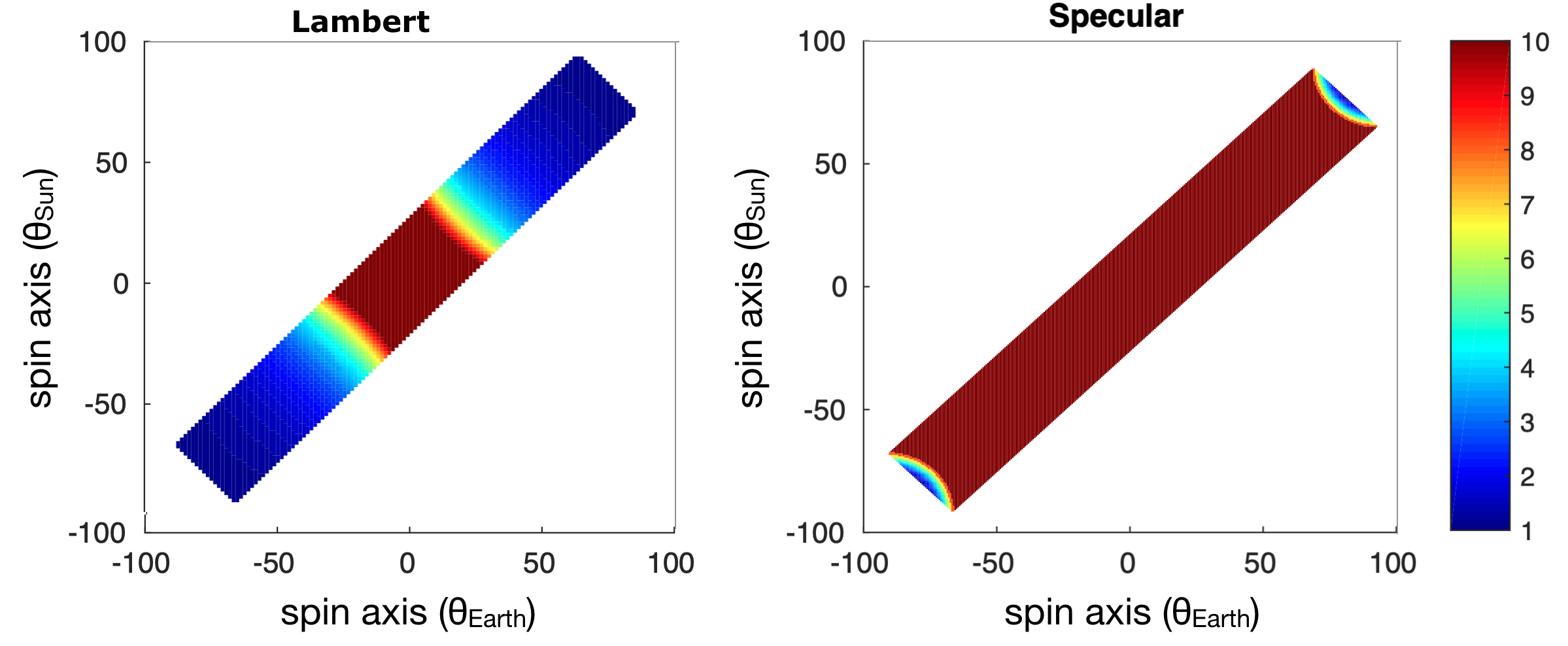}}
\caption{Color coded contours of the Lightcurve ratio from rotating ellipsoids of aspect ratio 6 as function of the spin axis angles for the four surface reflection properties: \bks (left), \ls surface (2nd), \lmb surface (3rd), and \spc reflection (right). In all case a fixed phase angle of $\Theta=24\degi$ is assumed. The probability to observe a lightcurve ratio higher than 10 when the object has aspect ration of 6 is about 3\% (\bks), 9\% (\ls), 30\% (\lmb) and 97\% (\spc). }\label{fig:prob3D}
\end{figure*}

The minimal aspect ratio calculated in the previous section would result in the observed lightcurve of \omu only if the spin axis orientation is exactly perpendicular. As this is unlikely, \omu must have a higher aspect ratio to be observed with such a lightcurve ratio.
Here, we calculate the probability to observe \omu lightcurve ratio, as a function of the object's aspect ratio assuming an isotropic spin orientation. 
In Fig.~\ref{fig:prob3D} we show the lightcurve ratio (color coded) for ellipsoids with aspect ratio of $a/b=6$. 
The lightcurve ratio is presented as a function of the angle between the spin axis and the Sun and the Earth orientation 
($\theta_\oplus$, $\theta_\odot$ in Eq.~\ref{eq:soe}).
We assume rotation only around the short (c) axis, from minimal energy consideration, and ignore tumbling (see model assumptions).
The white areas in the panels of Fig.~\ref{fig:prob3D} are excluded angles as they could not be realised with the phase angle of $\Theta=24\degi$.

To calculate the probability we give each point an appropriate weight according to an  isotropic distribution of the spin axis orientation, and a fixed phase angle.
The probability to observe a lightcurve ratio higher than 10 (dark red in Fig.~\ref{fig:prob3D}) when the object has aspect ratio of 6 is about 3\% (\bks), 9\% (\ls), 30\% (\lmb) and 97\% (\spc). 
For \spc reflection (right) the lightcurve ratio strongly depends on the aspect ratio (up to $(a/b)^4$ for perpendicular spin), and therefore the probability is high even for moderate aspect ratio - aspect ratio of 2 has probability of 53\% to observe a lightcurve ratio higher than 10.

For \spc reflection model, we are able to obtain this probability analytically.
A given lightcurve ratio $n$ can be observed by an angle $\phi_m$, satisfying
\begin{equation}
   \frac{F(\pi/2)}{F(\phi_m)} = n,
   \label{eq:lcr}
\end{equation}
where $F$ is given by Eq.~\ref{eq:spc2}.
The probability to observe light curve ratio larger than $n$ is now given by $\cos \phi_m$.
We find the probability to observe a lightcurve ratio of $n$ or larger from an ellipsoid with \spc surface reflection as a function of its aspect ratio $(a/b)$ to be:
\begin{equation}
    P= \frac{\sqrt{\left(a \over b\right)^2 - \sqrt{n}}}{\sqrt{\left(a \over b\right)^2 - 1}}
\end{equation}

\begin{figure}
\includegraphics[width=8.8cm]{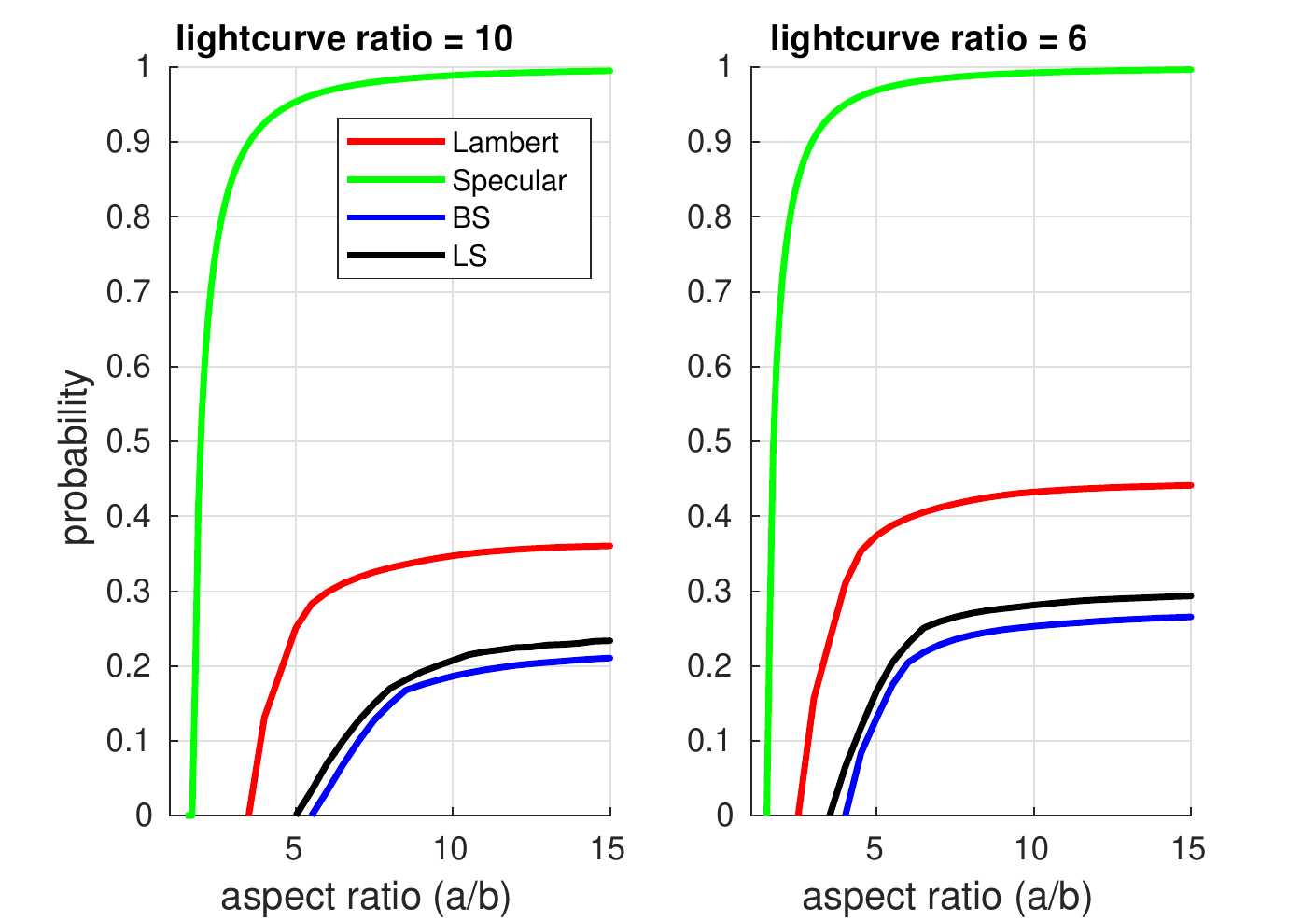}
\caption{The probability to observe a lightcurve ratio of minimum 10 (left) and 6 (right) as a function of the object aspect ratio, for different surface reflection conditions: \lmb (red), \spc (green), \bks (blue), and \ls (black). For all cases phase angle is $\Theta=24\degi$. }\label{fig:proball}
\end{figure}

In Fig.~\ref{fig:proball} we show probability to observe a light curve ratio larger than 10 (left) and 6 (right) as a function of the ellipsoid aspect ratio. All calculations are for a phase angle $\Theta=24\degi$.
As is shown in the figure, this probability increases with the aspect ratio up to a maximum probability. The increase depends on the model reflection method, namely how lightcurve ratio changes with aspect ratio (the slope in Fig.~\ref{fig:maxmin}).
For that reason the probability to observe a given lightcurve ratio is higher for \lmb surface object and much higher for the \spc surface, because of their stronger dependency of lightcurve ratio on aspect ratio.
However, the probabilities calculated here are not the likelihood to form such a surface, which might be lower for the \lmb and \spc models, based on the knowledge from our solar system.

\section{Discussion and conclusions}\label{cncl}

Since \omu is believed to be an interstellar object \citep{higuchi19} it might have different composition, age and history, than a solar system asteroid.
As a result, its surface may differ from what we know.
We show that the elongation of \omu can be much lower than predicted in previous models, by varying the surface reflection properties beyond the asteroid-like models.
The less elongated shape relieves the need for extreme albedo changes \citep{mashchenko19}, or unnatural origin \citep{bialy18}.

The question hence is what conditions can result in such surface properties. 
The surface properties of the rocky objects in the solar system are determined by the mineral crystallization as an outcome of the solar system composition and its formation.  
Different composition (as may be in other solar system), thermal history (heating), and dynamical history (friction, impacts) may result in different surface properties.
For example, processes that \omu may went through could melted it, like repeated flybys close to its star \citep{raymond18}, or heating by a red giant star \citep{katz18}.
If \omu is in addition a metal rich object, its current surface (after melting) may be glossy. However, there is a very small chance that a natural object has a perfect mirror surface. 
Space weathering during \omu interstellar travel could also changed its surface properties. In addition, the estimated age of \omu is less than 1\,Gyr \citep{almendafer18}, much less than small objects in the solar system.

The photometry data of \omu (reflectivity vs. wavelength) was compared to solar system spectral types. 
The spectral type of \omu is found to be close to D-type asteroids, Trojan asteroids, inner solar system populations, and small trans-Neptunian objects \citep{meech17,jewitt17,bannister17}. 
Still, error bars on the measurements are too large to determine whether it is similar to the solar-system objects \citep[e.g., Fig.\,5 in][]{jewitt17}. 
Moreover, spectral type does not give a unique characterisation of the surface properties.

From probability perspective, the extreme lightcurve that was observed for \omu indicates a spin axis orientation that is near perpendicular to the SOE plane. Small angles between the spin axis and the SOE plane cannot produce such a high lightcurve ratio, regardless of the object aspect ratio. Thus, the probability to get the observed lightcurve ratio is low, even for elongated bodies. As is shown in Fig.~\ref{fig:proball}, this maximum probability varies for the different surface models. As an outcome, the \lmb and \spc reflection models are found to allow for \omu observed lightcurve ratio with higher probability than the \bks and the \ls models. 
Yet, if our solar system is indicative, the formation of \lmb or \spc surface is less likely.

To summarise, as a first observed interstellar object, \omu introduced new challenges to small bodies theories. Some of the challenges are related to the irregular elongated shape that is derived from its lightcurve ratio. 
We show that if the surface reflection properties of \omu significantly differ from those of asteroids, its shape might be much less elongated. 
Those surface properties are also found to have higher probability to produce the observed lightcurve ratio of \omu, given that its spin axis orientation is unknown.
Although the models suggested here are for ideal ellipsoidal shape and ideal reflection method, the results emphasis the importance of surface properties for the derived aspect ratio.

In the coming years the interstellar object database is expected to grow \citep[e.g.,][]{trilling18}. 
Recently, a second interstellar object was detected, 2I/Borisov. 
Unlike \omu, 2I/Borisov is a comet-like object, with similar properties to solar system comets \citep[e.g.,][]{guzik19,jewitt19}.
Detection of more interstellar objects will help us understand their nature, and how different their surface properties are from what we know. 

\section*{acknowledgements}
We would like to thank the referee for the useful comments. We also like to thank Josef Durech, Karri Muinonen, and David Polishook for interesting comments and discussions. This research is partially supported by an Icore grant and an ISF grant.

\bibliographystyle{mnras.bst} 
\bibliography{allona.bib} 

\end{document}